# Low-cost Augmented Reality prototype for controlling network devices

Anh Nguyen*    Amy Banic[+]

University of Wyoming

**ABSTRACT**

With the evolution of mobile devices, and smart-phones in particular, comes the ability to create new experiences that enhance the way we see, interact, and manipulate objects, within the world that surrounds us. It is now possible to blend data from our senses and our devices in numerous ways that simply were not possible before using Augmented Reality technology. In a near future, when all of the office devices as well as your personal electronic gadgets are on a common wireless network, operating them using a universal remote controller would be possible. This paper presents an off-the-shelf, low-cost prototype that leverages the Augmented Reality technology to deliver a novel and interactive way of operating office network devices around using a mobile device. We believe this type of system may provide benefits to controlling multiple integrated devices and visualizing interconnectivity or utilizing visual elements to pass information from one device to another, or may be especially beneficial to control devices when interacting with them physically may be difficult or pose danger or harm.

**Keywords**: Augmented Reality, network device, mobile device.

**Index Terms**: Augmented Reality, Mobile Devices, Low-Cost, Off-The-Shelf, Bluetooth Network

## 1 INTRODUCTION

The growth of Augmented Reality (AR) technology in recent years has witnessed the emergence of numerous applications for educational, navigation, gaming and medical purposes [1]. In this paper, we present an AR prototype for controlling multiple devices, such as office networked devices. System administrators can operate network devices which are physically located in a distant area by remote logging into them. This method may be efficient when operating numerous devices in a data center but the operator person would not have the intuitive visual feedback of how devices would respond to a certain remote command or communicate with each other. AR technology is a potential solution to enhancing this experience.

Along with the development of electronic technology, office electronic devices, such as speakers, TVs, printers or cameras, are powered with more functionality. These devices can interconnect with each other via wireless communication. While all the surrounding devices are being developed with "smart" technology, it then becomes a challenge for the users to operate. Users need a more obvious and intuitive way to control the electronic devices surrounding them. AR technology has been adopted in attempts to deliver better experiences for users.

Hammond and Sharkey [2] have demonstrated the concept of integrating Augmented Reality with Home Systems using a head-tracker, a camera and a network device. Although delivering a great experience, this system may be too cumbersome for a home Do-It-Yourself setup. Another known problem of having multiple networked devices is that they are identified by IP addresses or MAC addresses only, therefore the remote controller can only produce a list of the devices but not their actual locations or other physical properties. To solve this problem, Mihara and Sakamoto [3] have proposed a method to operate multiple home network appliances, such as televisions, DVD records, and audio devices. All of these appliances are interconnected by the Digital Living Network Alliance, using a marker-based AR mobile application. In this system, an LED light serves as the AR maker, which is attached to an appliance in order to effectively specify its location under a low-light condition. Although this is an effective recognition solution, integrating the LED marker into every network device is not a low-cost and trivial solution.

While the previous work of Mihara and Sakomoto [3] focused on the marker-based AR recognition approach, we would like to propose an off-the-shelf, low-cost marker-less AR prototype for controlling network office devices. We also discuss various approaches to visualizing the information of the devices as well as the communication among devices (section 2 and 5) to provide a better controlling experience. Our system requires an Android mobile device to serve as the universal AR remote controller. The network devices to be operated would need to be able to support a wireless communication. The system promises to be a fun, interactive and low-cost way to remotely control network devices around your office such as printers, speakers, or laptops, or anywhere or remotely by way of webcam. Imagine that one can sit on a couch holding a smart-phone, scanning around the room to check the battery status of a charging wiimote or to control the volume of a speaker afar. Another use case is set in a laboratory environment, where there are many pieces of equipment. One can quickly and optionally remotely turn on or off, and modify various properties of multiple devices by one or two touches on their smart-phone. Users could more easily do this instead of operating each device manually. Additionally, another use case may be for devices which may be difficult to physically get to, such as those located in a difficult space or at an unreachable height, or where there is danger or harm, such as a use case set in a nuclear facility where it can be important to visually see the controllers but safer to interact with from a distance. This paper first describes our design work and then presents the initial technological prototype.

## 2 WHY A MOBILE AUGMENTED REALITY CONTROLLER

A mobile device such as a smart-phone could serve as a remote controller for network devices with a traditional user interface of menus, buttons, and sliders. However, this user interface is not intuitive and can be confusing to non-frequent users. In the problem of controlling network devices, we decided to use AR technology our prototype because AR provides users with added information that cannot be directly detected using their own senses [1].

One example is when a mobile device transmits music to a wireless speaker. In this case, the audio transmission can be visualized in an AR application instead of being displayed in a traditional menu-based interface:

*email: anguyen8@uwyo.edu
[+]email: abanic@cs.uwyo.edu

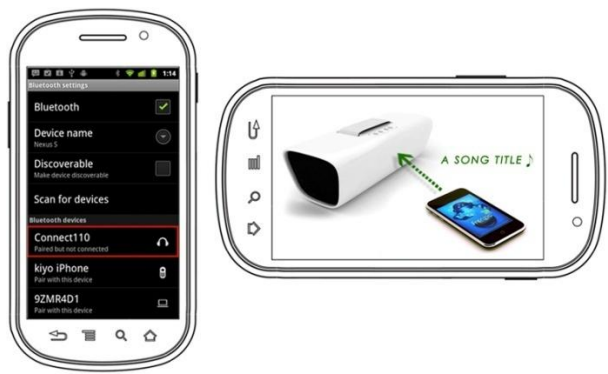

Figure 1: Traditional menu-based interface (left) and AR with visualization (right)

In figure 1, the menu-based interface may cause difficulty for users to identify the mobile device by only looking at its name. On the other hand, the visualization presents intuitively which the mobile device is currently transmitting music to the speaker, and what song is being played. Hence, the AR visualization is useful especially when there are many devices surround which causes difficulty for users to identify the communication between the devices. And this visualization of the communication between devices is a feature we propose in this prototype.

The visualization component in AR applications also shows to help users better in instructing them how to perform a manual task on an office device such as changing the toner cartridge of a printer. Metaio, one of the market leaders in the AR industry, released a sample AR application [9] that guides users step-by-step how to replace the toner cartridge of a printer by visualizing the internals of the printer and showing guiding arrows (Figure 2).Without this powerful AR application, users would normally have to read and follow the printer manual step-by-step. This is time-consuming and not very intuitive.

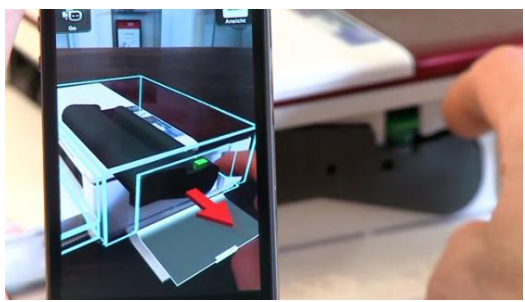

Figure 2: Visualization of a printer's internal system.

## 3 FRAMEWORK OVERVIEW

At this time, we are basing our prototype on a Bluetooth network as it is simple and quick to implement but it could be further replaced by a wider-ranged Wi-Fi network. Moreover, Bluetooth also enables a master device to communicate with a maximum of seven other slave devices. Our network (Figure 3) is setup in the way that the Android mobile device serves as the server (master) device pairing with each of the client (slave) device. The physical coverage range of a Bluetooth network for an Android smart-phone is about 8 meters [4] which is suitable for an office or home setup.

For the purpose of prototyping, we setup the framework with the following devices:
- A Galaxy Player 4.0with a built-in camera as the remote controller. The device runs Android 2.3.5.
- A HP ProBook 4530slaptop as the example of a slave Bluetooth device which requires additional software installed in order to be controlled.
- A Logitech Bluetooth speaker as the example of a Bluetooth device which does not require additional software.

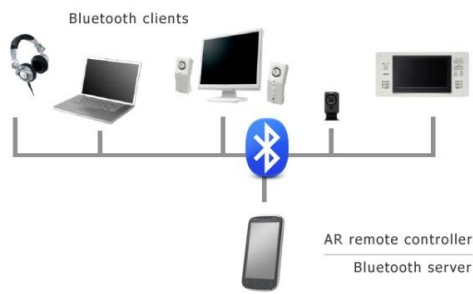

Figure 3: Network architecture

Our Android AR application on the Galaxy Player serves as the Bluetooth server and is able to send and receive data from the Bluetooth speaker and the laptop. To synchronize music and adjust the volume of the Bluetooth speaker, we only need to pair the Galaxy Player with the speaker, as the firmware on the speaker would be able to control it. However, in order to control the laptop such as controlling the mouse or adjusting its volume, it is necessary to install a small program on the laptop to receive and execute commands via Bluetooth. Currently, we have been able to send commands from the Android device to the laptop, but have not explored all the possibility of how to control the laptop. For the workshop demonstration, users will be able to remotely control both the speaker and the laptop.

## 4 IMPLEMENTATION

In order to rapidly implement the prototype, we chose to base our application on an open-source AR framework which is Vuforia [5]. The reason for choosing this framework is that it seamlessly supports Java Native Interface (JNI), which serves as the bridge between C++ code for the 3D rendering of virtual objects and Android Java-based user interface. We developed the virtual objects in our application using OpenGL 2.0 in C++. However, the downside of the framework is that we had to code in two languages and wire the communication between them via JNI. Another reason for choosing Vuforia is after experimenting with other AR open-source framework for Android, such as Metaio [6] and mixare [7], Vuforia produces the least amount of flickering on the screen when the camera is in motion.

Our AR application has three modes: registering, scanning, and controlling which are the main modules of the system.

### 4.1 Registering Bluetooth slave devices

The first and foremost functionality of our AR application is to register the slave devices so they can be recognizable by the system. This is possible by letting the application take a picture of the device (Figure 4). And this picture would help the program identify and distinguish one device from the others.

Other devices at your workstation can be registered in the same way (Figure 5). The requirement is that each picture has to have a good amount of details. The image recognition engine of Vuforia enables us to handle this registration.

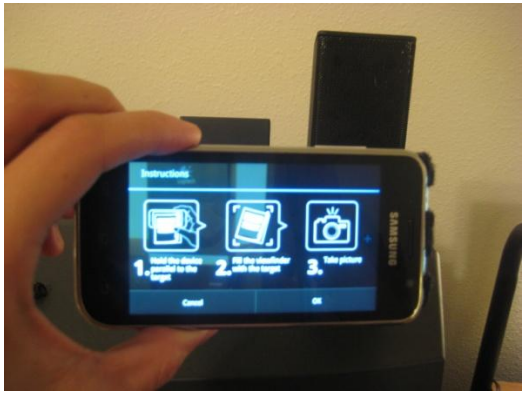

Figure 4: Registering our Bluetooth speaker

To increase the amount of details and therefore improve the device recognition rate, we could add a simple colored tag as the AR marker on each of the device. However, the prototype is intended for an office or laboratory setup where the light condition is sufficiently consistent; therefore we chose to implement this marker-less approach in which the devices are recognized and tracked by their registered images. Unfortunately, how Vuforia image recognition engine works specifically is a black-box to us because it is not open-source.

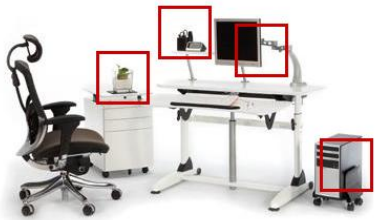

Figure 5: Registering multiple devices at your workstation

### 4.2 Device recognition and information retrieval

After the target devices have been registered, users can switch to scanning mode where the AR application tries to scan for the registered devices by matching the images in memory with the live streaming from the camera (Figure 6). A current limitation of Vuforia is to allow only five different devices to be recognized simultaneously.

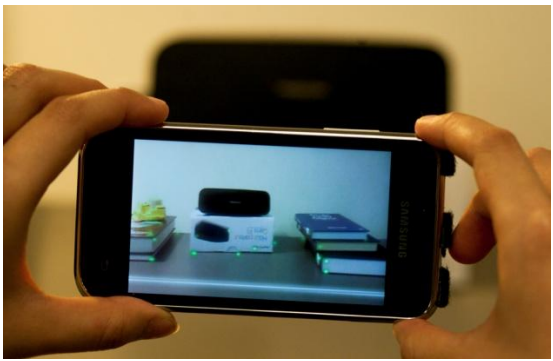

Figure 6: Scanning mode: trying to recognize the registered device

After a device has been recognized, a 3D overlay menu is displayed on top of the device showing the status of the device. Figure 7 shows the ID (*Memorex Bluetooth Speaker*), a power button, and a volume status (*75*) of the recognized Logitech speaker. Users can choose to adjust the volume slider or turn the speaker on or off.

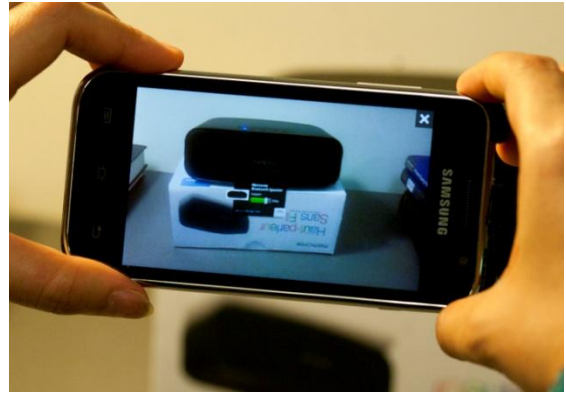

Figure 7: Displaying status of the recognized speaker

### 4.3 Operating recognized devices

We implemented a Bluetooth server application running on the Android device for continuous connection between the device and other clients such as the speaker. Users only need to pair the Galaxy Player with the speaker to be ready to operate it. The firmware of the speaker listens to commands over Bluetooth and operates the device accordingly.

Controlling the laptop is a bit more complicated as we would need to implement a little program that is able to carry out the OS-level operations such as controlling the mouse or logging off the computer depending on the need. We are currently developing a program in C for Linux that uses the open-source USB HID access library [8] to control the mouse according to the commands from the Galaxy Player via Bluetooth.

For better operating a device, beside the 3D overlay interface superimposed on the recognized device, we also provide users with the options to switch to a 2D interface when the device is afar and appears too small on the screen to interact with the controls on the 3D overlay. When the recognized device is not in the view, the 2D interface still allows the user to operate the device. In figure 9, the user is adjusting the volume when the speaker is not in the view anymore.

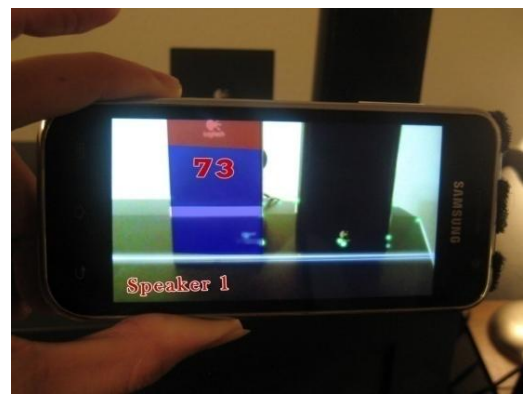

Figure 8: Visualization Overlays Actual Device

### 5 VISUALIZATION CAPABILITIES

Depending on the type of status that should be displayed to the user, a variety of visualization solutions could be utilized. In the case of determining the on/off status of devices, red and green (and optionally other colors for other types of status, such as

"stand-by") virtually outline the devices with a colored line or simple ON/OFF icons can display over the devices to allow for quick interpretation of status. In the case of adjusting volume for a device, a virtual slider (Figure 9) filled in with a particular color can serve multiple purposes: a) it demonstrates the amount of sound coming from the device, and b) provides direct manipulation for changing the level of sound through touch-and-drag interaction. In some cases, there may be an advantage to overlaying the visualization on the view of the device (Figure 8). Other possible visualizations may include data interconnectivity and virtual visual animated flow from one device to the other (Figure 1). A user could see documents being transferred to the printer, internet connectivity flow where high flow may indicate malicious activity, and slow to no connectively between devices will be quickly identified. Another possible visualization capability might be heat output of devices and temperature of room. A range of colors from one hue to another, can demonstrate temperature of devices and other sensor areas within the room, virtually overlaid on the devices, can help determine device proximity and adjustment of room temperature. There also will be benefits to allowing visualizations that take advantage of physical location memory where users can adjust properties of a device and the physical location of the virtual widget in relation to the device will assist the user in those adjustments. One final example would be to visualize documents or information and can use the AR application to visually drag and drop information from one device to the other. For example, dragging documents to the printer, documents from computer to large display or TV, images from a camera to a hard-drive and then to the printer, music from a media player to speakers, and so on.

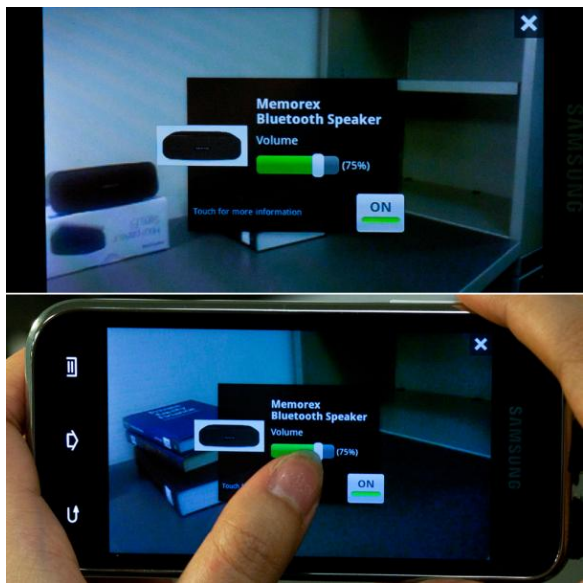

Figure 9: 2D overlay to better operate devices

## 6 DISCUSSION AND FUTURE WORK

This prototype is a work-in-progress and currently is able to register, recognize and operate the Bluetooth speaker properly via the AR application on a Galaxy Player. We plan to demonstrate these features at the workshop. The program to control the laptop is currently being developed as mentioned in section 3.3, and if it is finished on time, we would like to demonstrate it at the workshop as well.

*email: anguyen8@uwyo.edu
+email: abanic@cs.uwyo.edu

At this moment, our AR application can operate one device on the screen at a time; however, in the near future, we would like to make it recognize and operate two or more devices synchronously on the same screen. And with this capability implemented, we propose to visualize the communication between the network devices (Figure 1) which is transparent to the user. One possible foreseeable issue with synchronously augmenting multiple devices may involve the graphics rendering power limitation of the device.

We are experimenting with different approaches to designing AR controls including the 3D and 2D interfaces presented in section 4. A problem with designing for this particular AR application is that the screen size is too small for many controls to be placed on a screen. Especially when our prototype is used as a remote controller, the size of the slave devices afar on the screen would be relatively small. Hence, only three or less 3D overlay controls should be put on a screen for a device at a time, and if users need more advanced operations with many settings, we can provide an option to switch to the 2D interface or the traditional menu-based interface.

Another great future extension is to register a device using 6 images for 6 faces instead of one because that would increase the accuracy of object recognition dramatically and reduce the flickering problem when the camera is moving, thus generating more realistic augmented scenes. As a result, if scanned from any angle, the registered device could also be recognized.

## 7 CONCLUSION

This paper presents a low-cost, off-the-shelf, marker-less AR prototype which users can easily pair up their smart-phones with any other Bluetooth devices around and start operating them remotely. The prototype creates an interactive Augmented Reality experience for controlling devices around by turning your smart-phone into a remote controller. We find this application may be especially useful for a laboratory or an office environment where users have many network devices around and controlling them would not be a simple job. With this system users would have a universal remote controller for all the network devices.